# Permittivity model selection based on size and quantum-size effects in gold films.

Iuliia Riabenko*
*School of radiophysics, biomedical electronics and computer systems*
*V. N. Karazin Kharkiv National University*
Kharkiv, Ukraine
*Institute of Quantum Optics, Leibniz University Hannover, Welfengarten 1, 30167 Hannover, Germany*
jriabenko@karazin.ua ,
riabenko@iqo.uni-hannover.de

Sergey Shulga
*School of radiophysics, biomedical electronics and computer systems*
*V. N. Karazin Kharkiv National University*
Kharkiv, Ukraine
sergeyshulga@karazin.ua

Nikolai A. Makarovskii
*School of Physics*
*V. N. Karazin Kharkiv National University*
Kharkiv, Ukraine
nik.makarovsky@gmail.com

Konstantin Beloshenko
*School of radiophysics, biomedical electronics and computer systems*
*V. N. Karazin Kharkiv National University*
Kharkiv, Ukraine
kbeloshenko@karazin.ua

**Abstract:** The article explores optical properties of nanostructures containing spherical gold nanoparticles of various radii. We explore the particle radius as a criterion to select a permittivity model aimed at describing optical absorption spectra of gold granules. The experiments showed splitting of the absorption band of granular gold films to form a second absorption peak. The first peak is associated with the phenomenon of plasmon resonance, while the second one reflects quantum hybridization of energy levels in gold. Quantum effects were shown to prevail over size effects at a granule diameter of about 5-6 nm. The Mie theory gives a rigorous solution for the scattered electromagnetic field on a sphere taking into account optical properties of the latter, however, it does not specify criteria of selecting a model to calculate dielectric permittivity. Both calculations and experiments confirmed the limiting diameter of gold nanoparticles where the Hampe-Shklyarevsky model is applied. Meanwhile, this model was still unable to predict splitting of the plasma absorption band. The data presented in the article can be used for a predetermined local field enhancement in composite media consisting of a biolayer and metal nanoparticles.

1. **Introduction**

The Mie theory [1] is based on Maxwell equations which describe the fields where a plane monochromatic wave interacts with a spherical surface while physical properties of the latter differ significantly from those of the environment. The main difference lies in dielectric permittivity [2] which depends on the extinction coefficient [3], a parameter measured experimentally.

$$A = A_{abs} + A_{sct} = q\sigma_{ext} = q(\sigma_{abs} + \sigma_{sct}) \quad (1)$$

where $A$ consists of the absorption coefficient ($A_{abs}$) and the scattering coefficient ($A_{sct}$). It can also be presented as a product of $q$ and $\sigma_{ext}$ where $q$ denotes the particle filling factor (specific volume occupied by particles) and $\sigma_{ext}$ is the effective cross section of one particle extinction. The latter is obtained by adding the effective cross section of absorption by one particle ($\sigma_{abs}$) and the effective cross section of scattering by one particle ($\sigma_{sct}$).

The choice of the permittivity model for particles which scatter or absorb light determines the effective extinction cross section. However, it is more challenging to delineate the limits of applicability for a certain model describing permittivity of particles. When the particle radius is much smaller than the irradiation wavelength ($a \ll \lambda$), the formulas for the effective cross section of absorption and scattering are as follows [4]:

$$\sigma_{abs} = 4\pi \frac{\omega}{c} n_0 a^3 Im \frac{\varepsilon - \varepsilon_0}{\varepsilon + 2\varepsilon_0} \quad (2)$$

$$\sigma_{sct} = \frac{8\pi}{3} (\frac{\omega}{c} n_0)^4 a^6 \left|\frac{\varepsilon - \varepsilon_0}{\varepsilon + 2\varepsilon_0}\right|^2 \quad (3)$$

Where $\omega$ is the frequency of the electromagnetic radiation incident on the particle while $n_0$ refers to the refractive index of the medium surrounding the particle with the radius of $a$.

Thus, the effective extinction cross section determines the limits of a particular permittivity model for particles which either absorb or scatter light. For a continuous metal film, the Drude theory [5] is applied since it describes the reflection spectra well. This theory [5] describes permittivity as follows:

$$\varepsilon = 1 + \frac{\omega_p^2}{-\omega^2 + i\nu\omega} \quad (4)$$

where $\omega_p = \sqrt{\frac{4\pi n e^2}{m}}$ denotes the plasma frequency of free electrons.

This formula is based on the assumption that permittivity only relies on plasma of the free electron gas in the metal rather than size effects. One of the first attempts to record size effects was made by Maxwell Garnet [6] who suggested that metal spherical granules isolated from each other represent dipoles located on a non-absorbing dielectric substrate with $\varepsilon_0$ permittivity. In this case [7], the effective permittivity of the colloid $\varepsilon_{ep}$ is described by the following equation:

$$\varepsilon = (n - ik)^2 = 1 + \frac{3q \frac{n_m^2 - 1}{n_m^2 + 2}}{1 - q \frac{n_m^2 - 1}{n_m^2 + 2}} \quad (5)$$

where $n$ and $k$ refer to effective optical constants of the colloid while $n_m$ is the complex refractive index of the substance the granule consists of (assumed to be the same as that of metal).

The Drude theory is applicable to describe the reflection spectra of a continuous metal film [5], where only the free electron gas in the metal contributes to permittivity while size effects are not taken into account. The Maxwell Garnet model [6] used the Lorentz-Lorentz relation [8] to describe the effective permittivity of a colloid, which gives a result inapplicable in the case of a two-dimensional colloid of a granular film. On the other hand, the effective value of the complex dielectric constant of the film depends on the concentration of granules [9], not on their size.

Attempting to resolve the contradictions, Hampe [10] experimented with deposition of fine gold particles on fused quartz substrates. After annealing the prepared samples, Hampe used a spectrophotometer and an electron microscope for measurements to show that, despite the fact that annealing does not change the filling factor, the absorption maximum was increasing with a sharper peak as gold particle size was growing, and its position in the spectrum changed. Hampe also suggested that an anomalous absorption band is associated with oscillations of free electrons inside the granules. However, Hampe did not take into account the dipole-dipole interaction between the granules. All of the above ideas were combined in the Hampe-Shklyarevskii theory [11] where the values of the real and imaginary parts of the effective permittivity of a granular film were found to agree with the experiment [12]. This theory describes eigenfrequency as follows:

$$\omega_s^2 = \frac{\omega_p^2}{\varepsilon_m + 2\varepsilon_0} - \frac{\omega_p^2}{3(\varepsilon_m q + (1-q)\varepsilon_0)} \left(\frac{a}{b}\right)^3 S \quad (6)$$

The granule radius ($a$) and the distance between the granules ($b$) and $S$ as the sum that specifies the coordinates of each dipole granule relative to the origin were taken into account. However, the values obtained did not account for interzone transitions. However, the values obtained did not account for interzone transitions.

Thus, the hypothesis of a hybrid or combined nature of plasma resonance was put forward. Thus we divided $\varepsilon_1(\omega)$ [13] into two components, one of which corresponds to free electrons, and the other is related to interband absorption. The paper suggests that gold has zero values of $\varepsilon_1(\omega)$ at $\omega \sim 3.2$ eV. Thus, hybrid resonance can be associated with changes in motion of both d- and s-electrons. The presence of such a resonance must depend, in particular, on the oscillator strength and the frequencies of transitions under consideration. Thus, in [14], permittivity formulas were obtained for a colloid containing small metal spheres and optical properties of a colloid containing small silver particles were analyzed.

Based on the classical theory of dispersion, the complex dielectric constant can be represented as:

$$\varepsilon^!(\omega) = 1 + \sum_{i=1}^{n} \frac{\omega_{pi}^2}{\omega_{0i}^2 - \omega^2 + i\gamma_{0i}\omega} \quad (7)$$

where $\omega_{0i}$ – natural frequencies of the i-transition, $\gamma_{0i}$ – relaxation frequency, $\omega_{pi}^2 = \frac{4\pi n_i e^2}{m}$ – plasma frequency of the i-transition with the participation of $n_i$ electrons ($n_i = f_i n$ where $f_i$ – oscillator strength for the i-transition, $n$ – total concentration of electrons).

The dielectric constant of the considered environment can be found by adding up the optical properties of the spheres and their environment using the following expression [14]:

$$\varepsilon(\omega) = \varepsilon_0(1-q) + q \frac{\sum_i (\omega_{si}^2 \prod_{i \neq j} r_j)}{\prod_i r_i + \sum_i \omega_{si}^2 \prod_{i \neq j} r_j} \quad (8)$$

where $r_i = \omega_{0i}^2 - \omega^2 + i\gamma_{0i}\omega$ is the denominator in the i-term in (7) and $\omega_{si}^2 = \omega_{pi}^2 (1 + 2\varepsilon_0)^{-1}$. The products and the sum in (8) correspond to the number of oscillators. Based on this formula, the optical properties of a colloids can be calculated if the characteristics of individual oscillators $\omega_{0i}, \omega_{pi}, \gamma_{0i}$ are known.

2. **Material and Methods**

Samples were prepared for research according to the method described in detail in [15] with some improvements described below.

The Au film was sprayed onto a SiO$_2$ substrate in a $10^{-7}$ Torr vacuum by means of evaporation from a boat. The substrate was preliminarily cleaned in potassium dichromate and then treated with ion discharge under low vacuum conditions (approximately $10^{-2}$ Torr). The initial film thickness reached $100\ nm$. For the first time, a specifically modified laser implantation technology was used for thermal treatment of a gold film and partial implantation of gold nanoparticles into the near-surface layer of fused quartz in the most active band from a thermal point of view.

The gold film was exposed to thermal treatment using two lasers which were located on the same optical axis and were heating the sample with the film from both sides. A 10 W Holmium doped yttrium aluminum laser (Ho:YAG) with a wavelength of 2,1 $\mu m$ irradiated the gold film. The opposite side of the substrate was heated by a 25 $W$ $CO_2$ laser with a wavelength of 10,6 $\mu m$. The temperature gradient in SiO2 was controlled by a Fluke Ti400 infrared camera with a wavelength range of $8 - 15\ \mu m$. The Ho:YAG laser was chosen due to the fact that the emissivity of gold at a wavelength of 2,1 $\mu m$ is ~0,4 [16]. In other words, 40% of the laser radiation is absorbed by the gold film, while 60% is reflected. On the other hand, fused quartz absorbs light at a wavelength of $10,2 - 10,6\ \mu m$, which leads to e-fold attenuation of the laser radiation at the depth of about $1-2\ mm$. Since the substrates used in the experiment are 5 mms thick, the substrate is heated by the $CO_2$ laser beam at half the depth. Laser beams are at the same optical axis perpendicular to the substrate which gives an opportunity to heat the sample on both sides creating a local non-equilibrium state necessary to structure the gold film (Fig. 1).

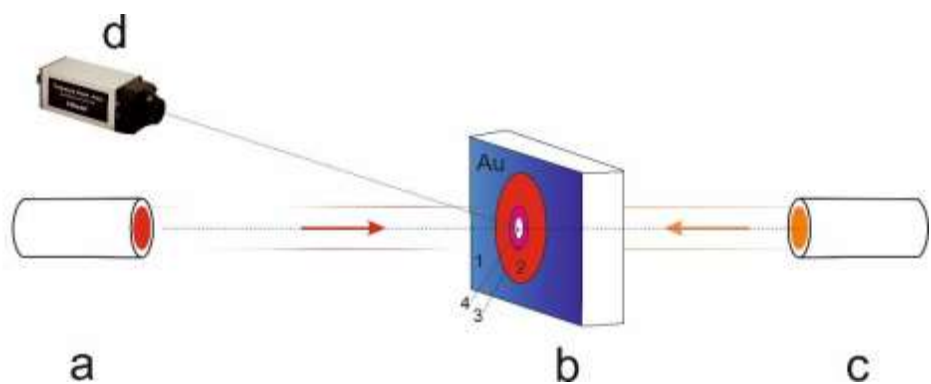

Fig.1 Experimental setup to form a nanoparticles cluster structure. a) neodymium-doped yttrium aluminum laser; b) substrate with a film (numbers indicate bands on the sample); c) $CO_2$ laser; d) Fluke Ti400 infrared camera

The temperature gradient was 384 $^0C$ per mm. After the exposure time of 15 minutes, pronounced color bands emerged on the surface of the sample in the transmission light. In band 1, the original film before irradiation was of a blue color. In bands 2, 3, 4, we observed a purple color under the temperature of 568 $^0C$, a pink color under 974 $^0C$, a dark gray under 1214 $^0C$, respectively. A Shimatzu UV2600 double-beam spectrophotometer was used to take spectrophotometric measurements for all the four bands. All quartz substrates, both reference samples and samples with applied films, were taken from the same batch, and the relief of the band surface was measured with an 801NER Pro AFM microscope.

3. **Results and Discussion**

After the exposure of the sample to laser radiation, we detected 4 bands of markedly different colors. The optical properties and dimensions of the film in the bands are shown in Table 1. We carried out spectrophotometry for all the bands (Fig. 2). The surface morphology of each band was examined with an AFM microscope (Fig. 3). The color of the bands depends on the filling factor q and the size of the cluster structure. It is also determined by whether gold was implanted into the near-surface layer of fused quartz.

Gold particles were implanted only in the central band which was confirmed by its chemical and mechanical resistance and stability of the optical properties after the chemical or mechanical impact.

**Table 1 The optical and dimensional properties of the film in the bands**

| Bands | Cluster size | Nanoparticle size | $\omega_s^{exp}$ | $\omega_s^{theor}$ |
|---|---|---|---|---|
| Blue | Continuous film | 50 nm | - | |
| Purple | 500 nm | 10-25 nm | $3.316\times10^{15}$ s$^{-1}$<br>568 nm | $3.25\times10^{15}$ s$^{-1}$<br>a~20 nm<br>b~150 nm<br>q=0.1 |
| Pink | 300 nm | 8-15 nm | $3.443\times10^{15}$ s$^{-1}$<br>547 nm | $3.4\times10^{15}$ s$^{-1}$<br>a~10 nm<br>b~30 nm<br>q=0.07 |
| Implantation band | - | 2-6 nm | $3.664\times10^{15}$ s$^{-1}$<br>514 nm<br>$4.956\times10^{15}$ s$^{-1}$<br>380 nm | $3.7\times10^{15}$ s$^{-1}$<br>a~3 nm<br>b~30 nm<br>q=0.008 |

The initial thickness of the film was calculated to be $100\ nm$, which corresponded to a continuous gold film. This fact was confirmed by photometric measurements: curve 1 in Fig. 2 is typical for a gold film. The model used to describe the permittivity is based on the Drude theory.

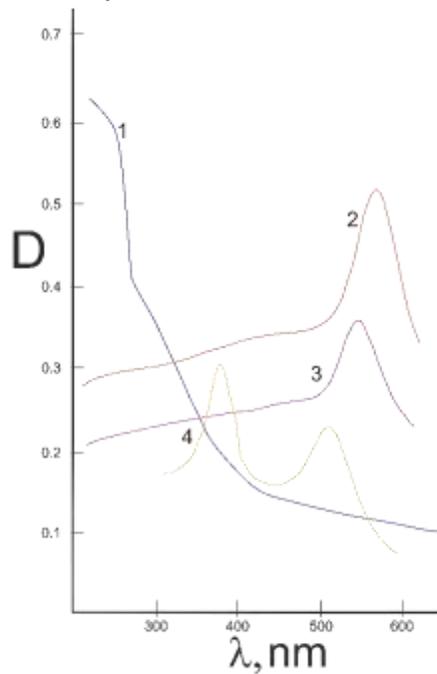

Fig.2 Optical density spectrum. 1 – continuous film; 2, 3 – irradiation band; 4 – implantation band.

When exposed to laser beams, the films undergo morphological changes which result in plasma resonance bands (curves 2, 3, 4 in Fig. 3).

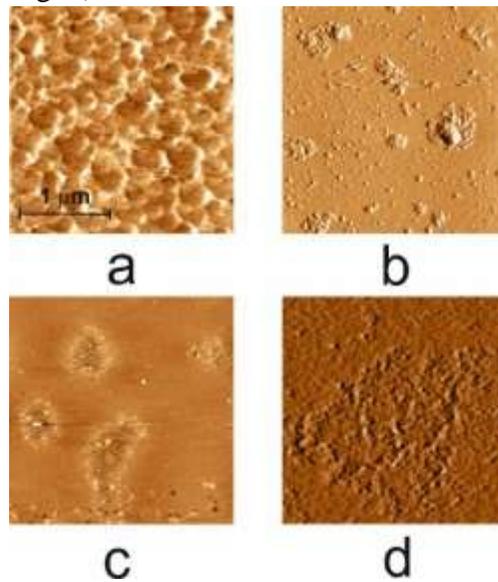

Fig. 3 AFM micrograph of the sample. a) blue band (polycrystalline gold film); b) purple band (cluster structure with large granules located separately); c) pink band (cluster structure with an increased distance between granules); d) implantation band ("crater" structure with granules implanted in the surface layer).

The AFM micrographs show the structural changes that the film undergoes in different thermodynamic bands. Plasma resonance bands and their positions indicate that a colloid is formed on the substrate surface while gold partially evaporates. The radius of the implanted nanoparticles is shown in Fig. 4 and is about $3\ nm$. Importantly, structural changes in the film lead to a shift in the plasma resonance band to an area with a short wavelength, which is well explained by the Hampe-Shklyarevskii theory. However, an additional absorption peak, not associated with plasma resonance, is detected in the central impact band. Instead, it is connected with interband transitions in gold granules. In this case the Hampe-Shklyarevskii theory cannot be applied to describe permittivity of metal granules. Quantum mechanics needs to be involved to describe this phenomenon.

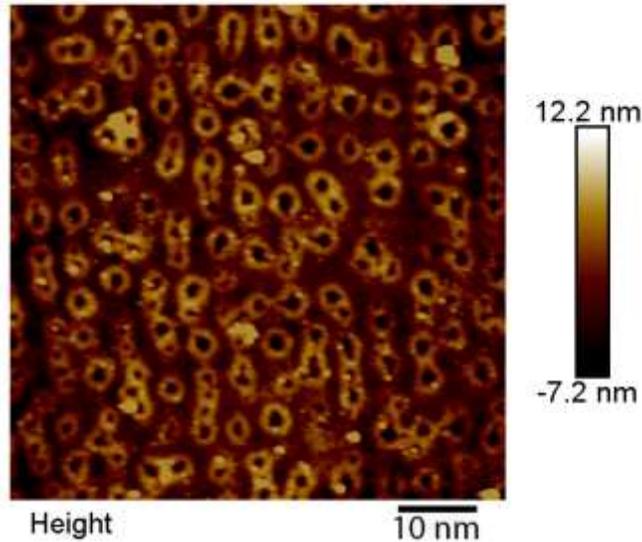

Fig. 4 AFM micrograph of the optical films implanted in the surface layer.

The experimental technique allowed us to simulate a case when islands with a diameter of about 150 $nm$ exposed to $CO_2$ radiation transformed into clusters with islands of a 30 nm diameter and with a distance between them equal to their diameter. Further irradiation led to the formation of colloids with a size of about 7 nm and a distance between clusters of approximately 150 nm. It gave an opportunity to study absorption spectra and surface morphology on the same substrate, which resulted in controllability and repeatability of the given parameters, such as the filling factor, the size of nanoparticles, and the distance between them. Plasma resonance frequencies for various parameters of the colloids are given in Table 1. Theoretical calculations were based on formula (6) used to determine the Frohlich frequency taking into account the dipole-dipole interaction between the granules. The permittivity of quartz $\varepsilon_0 = 2{,}15$ [9] and the permittivity of gold associated with interband transitions $\varepsilon_m = 2{,}13$ are referred to in [9]. The position of the second maximum cannot be described proceeding from the Hampe-Shklyarevskii model, however, its properties can be based on the energies of the energy levels in the metal granule described in the quantum theory of Sommerfeld [17].

The main objective of the research was to determine applicability of the Hampe-Shklyarevskii theory to describe permittivity of a granular colloid based on the size of granules. The fundamental assumption was the following: the Drude model is applicable if the nanoparticle size is approximately equal to the wavelength; in case of an isotropic colloid, when the distance between granules is much greater than the particle radius, it is the Maxwell Garnett model that can be applied; in case of clusters when this distance is approximately equal to the particle radius, the Hampe-Shklyarevskii model is appropriate. However, the limits of applicability of the latter depend on an increase in interband transitions.

In order to determine the size of nanoparticles on the basis of the permittivity model associated with inter-zone transitions, it can be assumed that the difference in the energies of the levels within the metal granule described in the Sommerfeld theory [17] should be approximately equal to the kinetic energy of the electron on the Fermi surface [18].

The eigenfrequencies of plasmons excited in small spheres can be determined with the following vector equation [14]:

$$Re\{Det|r_i\delta_{ij} \mp \omega_{sj}^2|\} = 0 \qquad (9)$$

In the presence of off-diagonal elements in (9), the eigenfrequencies of plasmons differ from $(\omega_{0i}^2 - \omega_{si}^2)^{1/2}$, which causes a hybrid resonance whose properties depend on the characteristics of all oscillators determining the optical changes in the environment. The understanding of such changes involves an approach based on quantum mechanics as well as the study of the electron energies in the metal. In this case, possible optical transitions on positive ions that form the core of a metal crystal are taken into account as well as the ionic component of the susceptibility $\chi_{ion}$ [19]. Thus, the impact of ionic susceptibility on optical transitions in gold can be explained as follows. Au has the $4f^{14}5d^{10}6s^1$ valence electron configuration [20]. When atoms condense to form a metal, electrons located in $6s^1$ orbitals turn into free electrons. In crystals, the outer $4f^{14}5d^{10}$ electrons form energy bands, and optical transitions from the $5d$ band to the $5s$ band partially filled with free electrons lead to an additional absorption band with a certain minimum threshold frequency $\omega_b$ [21]. For gold, $\omega_b$ lies in the visible area and, if $\omega_b > \omega$, the ionic susceptibility is a real value ($\chi_{ion}>0$). So, the eigenvalue of an electron in metal is determined by the following expression [22]:

$$E = \frac{2\pi^2 \hbar^2}{L^2 m}(l_1^2 + l_2^2 + l_3^2) \qquad (10)$$

where $l_1$, $l_2$, $l_2$ – quantum numbers, $m$ – electron mass, $L$ – length of the side of a cell, and $L > a$ – sides of an elementary cell in metal. The abovementioned mathematical equation can be written as follows:

$$\Delta E \sim m v_F^2 \qquad (11)$$

where $v_F = 1.8 \times 10^8 \; cm/s$ – the Fermi velocity (for gold [22])

By substituting (10) into (11) and performing the necessary calculations, we obtain

$$L = 2a \sim \sqrt{2}\frac{\pi\hbar}{mv_F}l \approx 5 \; nm \qquad (12)$$

These results agree with the values obtained with the AFM microscope, which showed that an increase in the absorption band associated with interband transitions is observed with a nanoparticle radius of about $3 \; nm$.

4. **Conclusion**

This paper covers the conditions which allow for applicability of different permittivity models used to describe optical spectra of composites containing gold nanoparticles. The applicability of different models was found to depend on the particle radius. We performed calculations and conducted an experiment to confirm the limiting diameter of gold nanoparticles where the Hampe-Shklyarevsky model is applicable. The article presents a methodology for determining the effects dielectric permittivity of metal nanoparticles is based on. Electrodynamic effects localized in space have generally been considered as dependent on the effects of plasmon-polariton interaction which accounts for the high-frequency peak of the absorption band, the latter being fully explained and well studied [9]. On the other hand, the effect associated with the interband absorption led to a shift of the plasma absorption band to the low-frequency region [23]. However, a separate absorption band in the low-frequency region, which would be clearly defined, has not been detected in previous studies.

The presented experiment revealed two spectrally separated bands. The high-frequency band shows plasmon-polariton interaction with the incident electromagnetic field, while the low-frequency band illustrates interband transitions in gold. Meanwhile, the experimental technique made it possible to create, on the one hand, films with a high volume filling factor for the granule material which is important for preserving the amount of absorbed energy by maintaining concentration of the substance; on the other hand, there are spatial regions in the system where quantum mechanical effects prevail. The Hampe-Shklyarevsky model does not predict for such effects. Thus, the implanted granule represents a pair of coupled oscillators [24]. One oscillator is a quantum dot whose resonant frequency coincides with the low-frequency peak, the other is a polariton whose resonant frequency corresponds to the high-frequency peak. The experimental data reveal how the near field affects the quantum mechanics system. Near field measurement techniques are well known in the microwave range [25, 26] but poorly developed in the optical range. The methodology presented in this article is intended to fill the existing gap. The conducted research provides a deeper understanding of the influence of a terahertz high-intensity electromagnetic field localized in the space on quantum dots. This technique is relevant for the study of surface-enhanced Raman scattering on a monomolecular biolayer. It is well known that the enhancement of Raman scattering is associated with high-intensity electromagnetic fields localized in the near field where the biolayer is located. No development of biosensors with predetermined metrological properties is impossible without a theoretical calculation of the electromagnetic field localization and experimental verification of the calculated data.

**Disclosures** The authors declare no conflicts of interest.
**Funding** The authors report there was no funding to declare.
**Data availability.** Data underlying the results presented in this paper are not publicly available at this time but may be obtained from the authors upon request.
During the study, all safety regulations were observed.